\documentclass[aps,prl,twocolumn,showpacs,superscriptaddress]{revtex4-1}

\usepackage{graphicx}  % needed for figures
\usepackage{dcolumn}   % needed for some tables
\usepackage{bm}        % for math
\usepackage{amssymb}   % for math
\usepackage[none]{hyphenat}  % avoid splitting of words at the end of lines
\usepackage{nicefrac}  % nice fractions within lines
\usepackage{amsmath}   % for math
\usepackage{footmisc}  % nice footnotes
\usepackage{latexsym}  % for widetext environment
\usepackage{romannum}  % roman numbering of sections
\usepackage{hyperref}  % for referencing 
\usepackage{multirow}  % for good tables 

\renewcommand{\arraystretch}{1.5}
\newcommand{\comment}[1]{}  % for commenting sections 

\newcommand{\Hc}{\mathcal{H}}
\newcommand{\kv}{\mathbf{k}}
\newcommand{\nb}{\bar{n}}

\newcommand{\Ev}{\mathbf{E}}

\begin{document}

\title{Quantum nonlinear mixing of thermal photons to surpass the
  blackbody limit}

\author{Chinmay Khandekar} \email{ckhandek@purdue.edu}
\address{Birck Nanotechnology Center, School of Electrical and
  Computer Engineering, College of Engineering, Purdue University,
  West Lafayette, Indiana 47907, USA}

\author{Liping Yang}
\address{Birck Nanotechnology Center, School of Electrical and
  Computer Engineering, College of Engineering, Purdue University,
  West Lafayette, Indiana 47907, USA}

\author{Alejandro W. Rodriguez}
\address{Department of Electrical Engineering, Princeton
  University, New Jersey 08544, USA}

\author{Zubin Jacob}
\address{Birck Nanotechnology Center, School of Electrical and
  Computer Engineering, College of Engineering, Purdue University,
  West Lafayette, Indiana 47907, USA}

\date{\today}

\begin{abstract}
Nearly all thermal radiation phenomena involving materials with linear
response can be accurately described via semi-classical theories of
light. Here, we go beyond these traditional paradigms to study a
\emph{nonlinear} system which, as we show, necessarily requires
quantum theory of damping. Specifically, we analyze thermal radiation
from a resonant system containing a $\chi^{(2)}$ nonlinear medium and
supporting resonances at frequencies $\omega_1$ and $\omega_2\approx
2\omega_1$, where both resonators are driven only by intrinsic thermal
fluctuations. Within our quantum formalism, we reveal new
possibilities for shaping the thermal radiation. We show that the
resonantly enhanced nonlinear interaction allows frequency-selective
enhancement of thermal emission through upconversion, surpassing the
well-known blackbody limits associated with linear
media. Surprisingly, we also find that the emitted thermal light
exhibits non-trivial statistics ($g^{(2)}(0) \neq 2$) and biphoton
intensity correlations (at two \emph{distinct} frequencies). We
highlight that these features can be observed in the near future by
heating a properly designed nonlinear system, without the need for any
external signal. Our work motivates new interdisciplinary inquiries
combining the fields of nonlinear photonics, quantum optics and
thermal science.
\end{abstract}

 \maketitle

\section{Introduction}

Nonlinear interactions between light fields are typically weak inside
bulk media but they can be significantly enhanced in resonant systems
to observe them at low optical powers. As the power requirements are
scaled down, the nonlinearities can influence not only quantum optical
processes~\cite{chang2014quantum} but also low power thermal radiation
phenomena (light fields generated by thermal fluctuations of
charges). The nonlinear mixing of thermal photons is a
theoretically-challenging, largely-unexplored topic which is relevant
in the field of thermal radiation with applications for renewable and
energy-conversion
technologies~\cite{fan2017thermal,tervo2018near}. Recently, it was
pointed out using semi-classical Langevin theory that resonantly
enhanced Kerr $\chi^{(3)}$ nonlinearities can be harnessed for
spectral-engineering thermal radiation~\cite{khandekar2015radiative,
  khandekar2015thermal} and near-field radiative heat
transport~\cite{khandekar2017near,khandekar2018near}. In that context,
the problem of thermal radiation from a \emph{passive} system of
coupled resonators of \emph{distinct} frequencies remained
unsolved. We solve it here using quantum theory for a system
containing a $\chi^{(2)}$ nonlinear medium and discover new
fundamental possibilities for the field of thermal radiation.

\begin{figure*}[t!]
  \centering \includegraphics[width=0.8\linewidth]{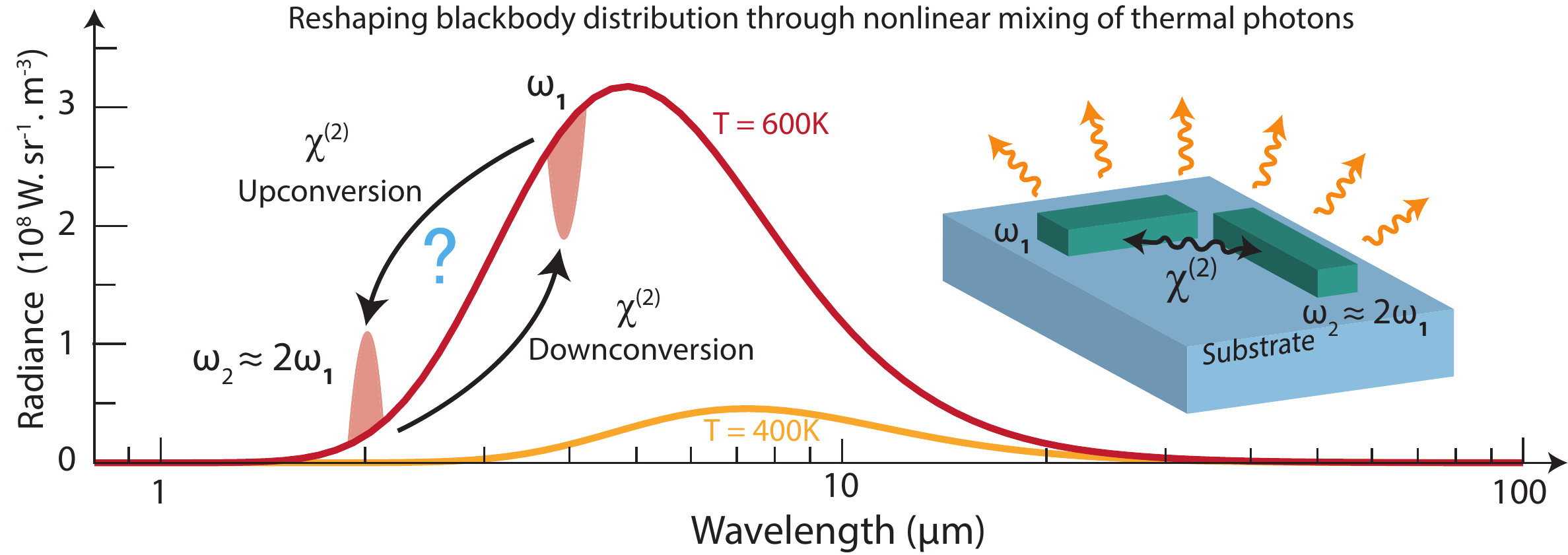}
  \caption{We rigorously analyze thermal radiation from a resonant
    system depicted in the inset. It is comprised of two resonators of
    frequencies $\omega_1$ and $\omega_2\approx 2\omega_1$, and
    contains a $\chi^{(2)}$ nonlinear medium. The resonantly enhanced
    nonlinear interaction will modify the Planck's blackbody
    distribution, potentially allowing enhancement of thermal emission
    at wavelengths where it is otherwise exponentially suppressed.  }
  \label{planck}
\end{figure*}

Our system depicted in figure~\ref{planck} consists of two resonators
at frequencies $\omega_1$ and $\omega_2 \approx 2\omega_1$, and
contains a $\chi^{(2)}$ nonlinear material. In stark contrast to prior
works in the quantum-optics literature where an external drive is
used~\cite{boyd2003nonlinear,Harris1990nonlinear,schmidt1996giant},
here the resonators are driven by low power thermal fluctuations
inside the medium. Because of $\chi^{(2)}$ nonlinearity, two photons
at $\omega_1$ can get upconverted to yield a single photon at
$\omega_2$, and a single photon at $\omega_2$ can get downconverted to
yield two photons at $\omega_1$. The resonantly enhanced, thermally
driven upconversion and downconversion processes can alter the
well-known Planck's distribution as illustrated in
figure~\ref{planck}, causing redistribution of thermal energy in
different parts of the frequency spectrum. 

%This has direct relevance
%for energy conversion applications~\cite{tervo2018near,
%  fan2017thermal,de2012conversion,luo2004thermal} and
%thermal-radiation engineering~\cite{li2018nanophotonic,
%  baranov2019nanophotonic}.

It turns out that the analysis of thermal radiation from such
nonlinearly coupled resonators of \emph{distinct} frequencies is not
trivial. Theoretically, it requires careful examination of their
thermal equilibrium behavior. It is known that the individual,
uncoupled resonators (harmonic oscillators) in equilibrium with a
reservoir at temperature $T$ contain an average thermal energy given
by the Planck's function
$\Theta(\omega,T)=\hbar\omega/[\textrm{exp}(\hbar\omega/k_B T)-1]$,
above the vacuum zero point energy~\cite{breuer2002theory}. In the
classical regime ($\hbar\omega \ll k_B T$), both oscillators have
equal average thermal energy ($k_B T$) by the equipartition law. If
they are nonlinearly coupled, we expect that thermally driven
upconversion and downconversion processes are balanced and thermal
equilibrium with the reservoir is maintained. However, in the
practically relevant nonclassical regime ($\hbar\omega \gtrsim k_B
T$), the oscillators have unequal average thermal energies and this
raises important fundamental questions. Is the balance between
upconversion and downconversion maintained? Are the coupled
oscillators in equilibrium with the reservoir? These questions must be
reliably answered by the theory before it is used to describe the
thermal radiation.
 
Almost all thermal radiation phenomena (in both classical and
nonclassical regimes) are described semi-classically using Kirchhoff's
laws~\cite{landau1980statistical,luo2004thermal}, fluctuational
electrodynamics~\cite{rytov1959theory,otey2014fluctuational} and
Langevin coupled mode
theory~\cite{zhu2013temporal,karalis2015temporal}. However, we find
that none of them can be readily extended for the problem at hand, and
the answers are found only within the quantum theory of
damping~\cite{breuer2002theory,scully1999quantum}. Within the quantum
formalism, the problem is non-trivial because of the lack of
closed-form analytic solutions and is unsolved in related early
works~\cite{agarwal1969quantum,drummond2014quantum,
  kozierowski1977quantum} . We solve it numerically using the
corresponding quantum master equation. Our analysis reveals that the
resonators are at thermal equilibrium with the reservoir along with
the balance between upconversion and downconversion processes. We also show the failure of alternative
semi-classical theory~\cite{khandekar2015radiative,dykman1975spectral}
which leads to unreasonably large deviation from thermal equilibrium. That comparison further
justifies the use of quantum theory for analyzing the nonlinearly
coupled passive resonators.

%We further show that an alternative semi-classical Langevin
%theory~\cite{khandekar2015radiative,dykman1975spectral} that otherwise
%accurately describes thermal fluctuations phenomena for singly
%resonant nonlinear oscillators even at non-classical frequencies
%($\hbar\omega \gtrsim k_B T$), leads to unreasonably large deviation
%from thermal equilibrium despite the frequency matching
%condition. This comparison indicates that the semi-classical theory
%cannot always be used for nonlinearly interacting resonators of
%distinct frequencies for which quantum theory becomes necessary.

We then use our quantum master equation approach to analyze the
thermal emission from the system. We show that by suitably engineering the system to favor the upconversion of thermal radiation from $\omega_1$
to $\omega_2$, one can enhance the thermal emission at $\omega_2$
beyond its linear thermal emission. For a resonant mode at $\omega_2$
of given linewidth, the maximum thermal emission using this nonlinear
mechanism is four times larger than the maximum thermal emission
achievable in the linear (Planckian) regime. The proposed mechanism
realizes the enhancement via spectral redistribution of thermal energy
which is fundamentally different from conventional approaches limited
to linear media for enhancing far-field~\cite{biehs2016revisiting,
  fernandez2018super,thompson2018hundred} and
near-field~\cite{yang2018observing,kim2015radiative} thermal
radiation. We further note that, to guide the design of
energy-conversion technologies~\cite{fan2017thermal,tervo2018near},
recent interesting
works~\cite{miller2016fundamental,buddhiraju2018thermodynamic,
  molesky2019bounds,molesky2019fundamental} have explored bounds on
radiative heat transfer in the linear regime. In that context, the
finding that the linear bound can be surpassed by the nonlinear
upconversion of thermal energy is new and important. Recently, the
far-field thermal emission from finite-size subwavelength objects is
termed as super-Planckian because it greatly exceeds the flux emitted
by a blackbody of the same geometric area. However, since Kirchhoff's
law is still valid and there is no enhancement when the absorption
cross-section is used instead of the geometric cross-section, it
remains an open question whether this geometric enhancement can be
identified as super-Planckian. In the context of this interesting
topic~\cite{biehs2016revisiting,fernandez2018super,
  thompson2018hundred,greffet2018light}, we note that Kirchhoff's law
is not applicable in the present work and the mechanism of nonlinear
upconversion paves the way for a regime beyond the super-Planckian
enhancement.

%also invites new inquiries of generalized Kirchhoff's %laws~\cite{miller2017universal} and thermal-emission sum rules for %nonlinear media.

Apart from the thermal-emission enhancement, we also find other
surprising effects. First, the autocorrelation $g^{(2)}(0)$ of emitted
thermal light deviates noticeably from the known value of
$g^{(2)}(0)=2$ for thermal light. And second, the emitted thermal
radiation exhibits strong correlations between the intensities
collected at two distinct frequencies of $\omega_1$ and
$\omega_2$. These features are surprising since they are observed by
heating the nonlinear system, without using any external signal. From
the perspective of quantum optics, we further note that our analysis
is markedly different from the analysis of the same system driven by
an external coherent
pump~\cite{agarwal1969quantum,drummond2014quantum,kozierowski1977quantum}.
Most notably, its importance for spectral-engineering thermal
radiation~\cite{li2018nanophotonic,baranov2019nanophotonic} is not
explored. The nascent topic of thermal-radiation engineering using
nonlinear media~\cite{khandekar2015radiative,khandekar2015thermal,
  khandekar2017near,khandekar2018near,soo2018fluctuational} remains to
be explored experimentally. However, we are confident that recent
theoretical inquiries including this work will motivate the
experiments in the future because of their fundamentally advancing
nature and practical utility.

%Since thermal-radiation engineering is primarily concerned
%with the control of magnitude, directionality, spectrum and
%polarization~\cite{li2018nanophotonic,baranov2019nanophotonic}, these
%features indicate new possibilities of tailoring the statistics and
%frequency correlations by heating the nonlinear system, without using
%any external signal.

%because of orders of magnitude difference in the number of photons and
%other technical details of the pump signal.

%In context of other recent
%works~\cite{lu2016biphoton} exploring related features in
%\emph{driven} systems, here we show that they can be observed by
%heating the nonlinear system without using any drive.

%All these effects are detectable by measuring the
%intensities in an actual experiment and provide new degrees of freedom
%for shaping the far-field thermal
%radiation~\cite{li2018nanophotonic,baranov2019nanophotonic}. 

%We note that the nonlinearities in context of other damped quantum
%harmonic oscillators have been
%studied~\cite{dykman2012fluctuating,haake1986master,
%  milburn1986dissipative,daniel1989destruction, peinova1990exact}, but
%the specific problem of thermal equilibrium of coupled oscillators of
%distinct frequencies has not been formally addressed to the best of
%our knowledge. 

\section{Quantum theory}
\label{sec:quantum}

We consider a system of two photonic resonant modes of frequencies
$\omega_j$ described by operators $\{a_j,a_j^{\dagger}\}$ for
$j=[1,2]$ and having frequencies $\omega_1,\omega_2 \approx
2\omega_1$. The resonators contain a $\chi^{(2)}$ nonlinear medium
which facilitates coupling between the resonators via upconversion and
downconversion processes. The system Hamiltonian
is~\cite{agarwal1969quantum,drummond2014quantum}:
\begin{align}
  \Hc_{\text{sys}} &= \hbar\omega_1 a_1^\dagger a_1 + \hbar\omega_2
  a_2^\dagger a_2 + \hbar(\kappa a_1^{\dagger^2}a_2 + \kappa^* a_1^2
  a_2^\dagger)
  \label{Hsys}
\end{align}
where $\kappa$ denotes weak nonlinear coupling ($\kappa \ll \omega$)
between the modes. The connection of the above phenomenological
Hamiltonian with Maxwell's equations can be made by comparing with
corresponding classical coupled mode theory widely used in
nanophotonics and derived using Maxwell's
equations~\cite{joannopoulos2011photonic}. The resonant frequencies
are calculated by solving the eigenmodes of Maxwell's equations while
decay and other coupling rates are obtained using perturbation
theory~\cite{bravo2007modeling,rodriguez2007chi}. Using this approach,
we show in the appendix that the nonlinear coupling $\kappa$ depends
on the overlap integral of linear resonant mode profiles.

The resonators are further coupled linearly with an intrinsic
(dissipative) environment/heat bath of harmonic oscillators described
by $\{b_{\kv,d},b_{\kv,d}^\dagger\}$. The radiation from the
resonators into an external environment or other channels is modeled
by linear coupling with radiation modes which are also harmonic
oscillators described by operators
$\{b_{\kv,e},b_{\kv,e}^\dagger\}$. The total Hamiltonian is:
\begin{align}
\Hc = \Hc_{\text{sys}} +\sum_{\substack{\kv, l=[d,e] \\ j=[1,2]}}
\hbar g_{\kv,l,j}[b_{\kv,l}^\dagger a_j+ b_{\kv,l}a_j^\dagger] +
\hbar\omega_{\kv} b_{\kv,l}^\dagger b_{\kv,l}
\label{Htotal}
\end{align}
Using the standard techniques of analyzing open quantum
systems~\cite{breuer2002theory,scully1999quantum}, we solve the
reduced dynamics of the two oscillators by tracing over the heat bath
degrees of freedom. In the regime of weak coupling $g_{\kv,l,j}$ with
a continuum of bath oscillators, we assume that the initial state is
$\rho \otimes \rho_{B}$ where $\rho$ is a joint density operator of
coupled resonators and $\rho_B$ is an equilibrium density operator of
bath oscillators. Using Markov approximation and Gibbs distribution
for bath oscillators, we obtain the following simplified master
equation in the Schrodinger picture:
\begin{align}
\dot{\rho} &= -i[\Hc_{\text{sys}}, \rho] \nonumber \\ &+
\sum_{\substack{l=[d,e] \\ j=[1,2]}} \bigg[\gamma_{j,l}
  (\nb_{\omega_j,T_l}+1)[2a_j \rho a_j^\dagger - a_j^\dagger a_j\rho -
    \rho a_j^\dagger a_j] \nonumber \\ &\hspace{35pt} +
  \gamma_{j,l}\nb_{\omega_j,T_l}[2a_j^\dagger \rho a_j -
    a_ja_j^\dagger\rho - \rho a_j a_j^\dagger] \bigg]
\label{master}
\end{align}
The above well-known form captures the interaction of each resonator
with heat bath/environment using the associated decay rates
$\gamma_{j,l}$. Here, $j=[1,2]$ denotes the resonator mode and
$l=[d,e]$ denotes either dissipative ($d$) heat bath or extrinsic
($e$) radiative environment at temperature $T_l$. For a practical
system, the decay rate is related to the quality factor $Q$ of the
resonator via the relation $\gamma =\omega/2Q$. Also, 
$\nb_{\omega,T}=1/[\mathrm{exp}(\hbar\omega/k_BT)-1]$ is the mean
photon number of a harmonic oscillator of frequency $\omega$ at
thermodynamic equilibrium temperature $T$.

%We note that the nonlinear
%effects at few photon numbers or low power thermal noise in absence of
%any pump are known to be so weak that the regime $\kappa \sim
%\omega_j$ is practically beyond the reach~\cite{chang2014quantum} and
%therefore, it will not be considered here.

%Because of this perturbative
%nature of optical nonlinearity, we need not transform the oscillators
%into dressed modes, which has been explored for damping of strongly
%but \emph{linearly} coupled composite
%systems~\cite{chou2008exact,joshi2014markovian}.

It is straightforward to obtain the temporal dynamics of important
relevant quantities from the master equation (Eq.~\ref{master}) such as
mean photon numbers $\langle a_j^\dagger a_j \rangle$ for $j=[1,2]$,
where $\langle \cdots \rangle$ denotes the quantum statistical
average.
\begin{align}
  \label{a1pa1m}
\frac{d}{dt}\langle a_1^\dagger a_1 \rangle &= -2[i\kappa \langle
  a_1^{\dagger^2}a_2 \rangle - i\kappa^* \langle a_1^2 a_2^\dagger
  \rangle]\nonumber \\ &\hspace{20pt} -\sum_{l=[d,e]}
2\gamma_{1,l}[\langle a_1^\dagger a_1 \rangle - \nb_{\omega_1,T_l}]
\\ \frac{d}{dt}\langle a_2^\dagger a_2 \rangle &= [i\kappa \langle
  a_1^{\dagger^2}a_2 \rangle - i\kappa^* \langle a_1^2 a_2^\dagger
  \rangle ]\nonumber \\ &\hspace{20pt}-\sum_{l=[d,e]}
2\gamma_{2,l}[\langle a_2^\dagger a_2 \rangle - \nb_{\omega_2,T_l}]
\label{a2pa2m}
\end{align}
From these equations, one can obtain the rates of energy loss for each
mode by multiplying Eq.~\ref{a1pa1m} by $\hbar\omega_1$ and
Eq.~\ref{a2pa2m} by $\hbar\omega_2$. In the steady state, it then
follows that the total power flow from mode $j=[1,2]$ to bath
$l=[d,e]$ is:
\begin{align}
P_{j \rightarrow l} = 2\gamma_{j,l}\hbar\omega_j[\langle
  a_j^\dagger a_j \rangle - \nb_{\omega_j,T_l}]
\end{align}

%It includes the power lost by mode $j$ to bath $l$ given by
%$2\gamma_{j,l}\hbar\omega_j\langle a_j^\dagger a_j \rangle$ as well as
%the power gained by mode $j$ from bath $l$ given by
%$2\gamma_{j,l}\hbar\omega_{j}\nb_{\omega_j,T_l}$.

Similarly, the total
power flows leaving the modes via nonlinear coupling $\kappa$ are:
\begin{align}
  \label{P12}
P_{1 \rightarrow 2} &= 2\hbar\omega_1[i\kappa \langle
  a_1^{\dagger^2}a_2 \rangle - i\kappa^* \langle a_1^2 a_2^\dagger
  \rangle] \\ P_{2 \rightarrow 1} &= -\hbar\omega_2[i\kappa \langle
  a_1^{\dagger^2}a_2 \rangle - i\kappa^* \langle a_1^2 a_2^\dagger
  \rangle] \label{P21}
\end{align}
The equality $P_{2 \rightarrow 1}= -P_{1\rightarrow 2}$ when $\omega_2
=2\omega_1$ indicates the energy conservation satisfied by the
underlying microscopic nonlinear process. Deviation from this
frequency matching condition has important implications for
equilibrium mean photon numbers of the oscillators as we explain
further below.

The above expressions also clarify the measurable quantities in an
actual experiment. In a practical system, the radiative heat
transferred to the external environment at a lower temperature $T_e <
T_d$ can be measured by a suitable detector. By collecting the terms
in Eqs.~\ref{a1pa1m} and \ref{a2pa2m} corresponding to the energy
exchange with the external heat bath, we can quantify the far-field
thermal emission power from the resonators as,
\begin{align}
P_{j}^{\text{far-field}}=2\gamma_{j,e}\hbar\omega_j[\langle
  a_j^\dagger a_j \rangle - \nb_{\omega_j,T_e}]
\label{power}
\end{align}
for $j=[1,2]$. The calculation of these relevant quantities in the
nonlinear regime ($\kappa \neq 0$) is however not as straightforward
as it is for linear systems. Although quantum Langevin equations for
the operators can be written in the nonlinear regime, they cannot be
analytically solved. Also, the equations of motion for operators such
as $\langle a_1^{\dagger^2}a_2 \rangle $ obtained from the master
equation contain higher order terms such as $\langle
a_1^{\dagger^3}a_1a_2 \rangle$, leading to an infinite set of
equations where higher order terms are not necessarily
negligible. Therefore, we directly obtain the steady state density
matrix from the master equation by numerically solving
it~\cite{yang2019engineering,yang2019quantum}. We write the master
equation (\ref{master}) in its matrix form in the basis of photon
number states. The vector form of the joint density operator is (see
Appendix B in~\cite{yang2019engineering})
\begin{align}
  |\rho) =\sum_{\substack{n_1=0\\m_1=0}}^{\infty}
  \sum_{\substack{n_2=0 \\ m_2=0}}^{\infty} \rho_{n_1,m_1; n_2,m_2}
  |n_1,m_1; n_2,m_2)
\label{eqrho}  
\end{align}
with $|n_1,m_1; n_2,m_2) = (|n_1\rangle \otimes \langle n_2|) \otimes
(|m_1\rangle \otimes \langle m_2|)$. Here the notation $|v)$ is
introduced to denote a vector in Liouville space which is not a ket
state or a matrix. $n$ denotes number of photons at $\omega_1$ and $m$
denotes number of photons at $\omega_2$.  It follows from the trace
condition ($\text{Tr}\rho =1$) that $\sum_{n}\sum_{m} \rho_{n,m;
  n,m}=1$. We obtain the steady state solution of the density matrix
by solving the master equation (\ref{master}) and the trace condition
together. We work in the regime where this problem can be solved
numerically by introducing a cutoff $n_c$ limiting the upper bounds in
\eqref{eqrho} such that all the terms $\rho_{n_1,m_1; n_2,m_2}$ beyond
the cut-off ($n_j,m_j >n_c$) are negligible and can be safely
ignored. We estimate the cutoff using the known linear solution
(uncoupled oscillators) such that the probability of occupation of
photon number eigenstates above the cutoff is exponentially smaller
than 1. While these cut-offs differ for oscillators of different
frequencies, we choose a cut-off larger than the maximum of the
cutoffs for individual oscillators. We also verify that the numerical
solution is well-converged with respect to $n_c$.

We note that the nonlinear interaction term in the system Hamiltonian
given by Eq.~\ref{Hsys} commutes with the Hamiltonian of uncoupled
oscillators under the frequency matching condition
($\omega_2=2\omega_1$). Therefore, despite the nonlinear coupling, the
above solution using the photon number eigenstates is justified. For
small mismatch in the frequencies ($\omega_2\approx 2\omega_1$), the
solution is approximate. However, we argue that it is reasonable if
the frequency mismatch is of the order of nonlinear coupling
($\kappa$). For practical systems, the nonlinearity is very weak
($\kappa \lesssim 10^{-5}\omega_1$ described further below) and for
strong nonlinear effects on thermal radiation, the condition $\kappa
\sim \gamma$ necessitates the use of resonators of similar linewidths
($\gamma \sim \kappa$). We find that a small frequency mismatch
($|\omega_2-2\omega_1| \sim (\kappa,\gamma)$) affects the relevant
physical quantities with relative deviations of
$\mathcal{O}\{\kappa/\omega\} \lesssim 10^{-5}$ which can be
considered negligible for practical purpose.

\section{Thermal equilibrium of resonators}

We first analyze the situation where the local resonator heat bath
temperature $T_d$ is the same as the temperature of the external
environment $T_e$. In the absence of nonlinear coupling $\kappa$, it
is straightforward to obtain the steady state density matrix
$\rho_{n_1,m_1; n_2,m_2}$
analytically~\cite{breuer2002theory,scully1999quantum}. In the photon
number basis, it is diagonal ($\rho_{n,m;n,m}$) and is given below:
\begin{align}
  \rho_{n,m;n,m} = \frac{\nb_{\omega_1,T}^n
    \nb_{\omega_2,T}^m}{(\nb_{\omega_1,T}+1)^{n+1}
    (\nb_{\omega_2,T}+1)^{m+1}}
  \label{linsolution}
\end{align}
In the presence of nonlinear coupling $\kappa$, by employing our
numerical approach, we find that the same steady state solution (with
zero off-diagonal elements) satisfies the master
equation~(\ref{master}) when the frequency matching condition
$\omega_2=2\omega_1$ is strictly satisfied. The linear solution leads
to zero nonlinear correction to the energy of the resonant system,
$\text{Tr}[\rho(\kappa a_1^{\dagger^2}a_2 + \kappa^* a_1^2
  a_2^{\dagger})]=0$, despite the nonlinear coupling ($\kappa \neq
0$). Consequently, the nonlinear terms in \eqref{a1pa1m} and
\eqref{a2pa2m} vanish leading to thermal equilibrium with bath
oscillators in the strictest sense ($\langle a_j^\dagger a_j \rangle
=\nb_{\omega_j,T}$).

\begin{figure}[t!]
  \centering \includegraphics[width=\linewidth]{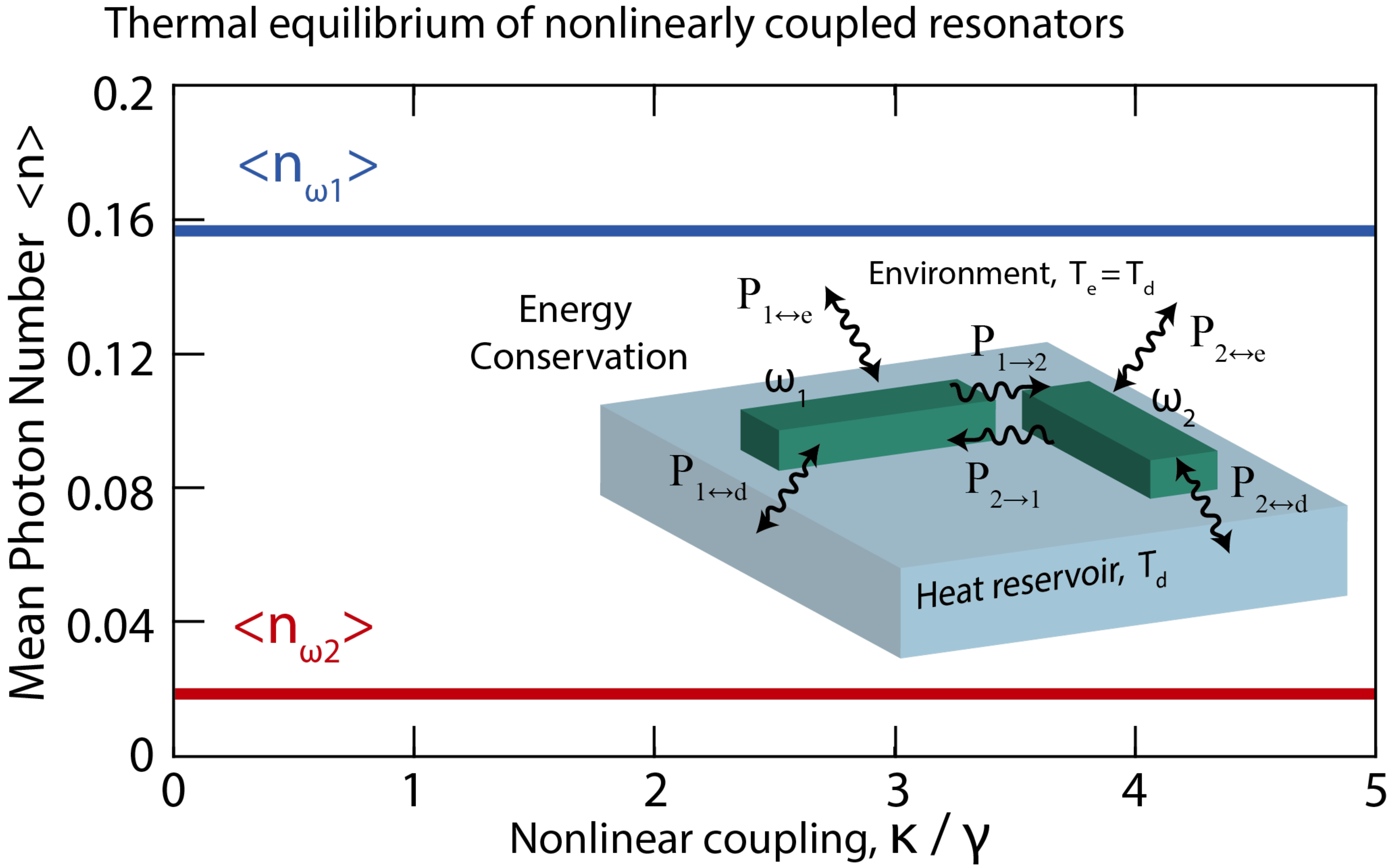}
  \caption{Steady state mean photon numbers ($\langle n_{\omega_j}
    \rangle$) of nonlinearly coupled resonators ($\kappa \neq 0$) are
    equal to equilibrium mean photon numbers ($\nb_{\omega_j,T}$) of
    bath oscillators ($T_d =T_e =T$) irrespective of nonlinear
    coupling, decay rates and temperatures (higher or lower mean
    photon numbers). The steady state density matrix is diagonal and
    same as in the linear regime given by Eq.~\ref{linsolution}. The
    inset demonstrates all energy flux rates that balance each other
    at thermal equilibrium.}
  \label{figeq}
\end{figure}

Figure~\ref{figeq}(a) provides an illustration by considering resonant
frequencies as $\omega_1=1.57\times 10^{14}$rad/s ($\sim 12\mu$m),
$\omega_2=2\omega_1$ ($\sim 6\mu$m) and temperatures
$T_d=T_e=T=600$K. The schematic shows the system and various power
flows. The mean photon numbers of bath oscillators at these
frequencies are $\nb_{\omega_1,T}=0.159$ and $\nb_{\omega_2,T}=0.02$.
As demonstrated in this figure, despite the finite nonlinear coupling
($\kappa \neq 0$), the resonator mean photon numbers are equal to
those of bath oscillators ($\langle n_{\omega_j} \rangle=\langle
a_j^\dagger a_j \rangle$).  This equality holds irrespective of the
temperatures $T$ (higher or lower mean photon numbers of bath
oscillators), the coupling parameter $\kappa/\gamma$, and the decay
rates $\gamma_{jd}/\gamma$ and $\gamma_{je}/\gamma$. We have
normalized these parameters with a suitable $\gamma$ which
characterizes the typical linewidth of the resonators. While the
nonlinearity is perturbative in comparison with frequencies ($\kappa
\ll \omega_j$), we note that the above results are obtained in the
strong nonlinear regime ($\kappa \sim \gamma$). As we show in the next
section, strong nonlinear effects are indeed observed for the same
parameters but under nonequilibrium condition ($T_d \neq T_e$) because
of subtle modifications of power flows that otherwise balance each
other when the reservoirs are at thermal equilibrium ($T_d = T_e$).

We note that other nonlinear processes of the form
$\omega_1+\omega_1'=\omega_2$ where $\omega_1\neq \omega_1'$ are also
possible in the presence of $\chi^{(2)}$ nonlinearity. But, here we
focus on extensively studied second-harmonic generation process
($\omega_2=2\omega_1$) from the perspective of potential
experiments. If the frequencies are mismatched slightly
($|\omega_2-2\omega_1|\sim \gamma,\kappa$), the above solution using
the photon number eigenstates is approximate as we noted
earlier. While very small relative deviations may arise because of the
approximation, it can be argued that the oscillators remain at thermal
equilibrium with the heat bath. In particular, the nonlinear energy
exchange rates given by Eqs.~\ref{P12} and~\ref{P21} oscillate
sinusoidally over a timescale $\Delta t \sim 1/|\omega_2-2\omega_1|$
in the absence of frequency matching which becomes evident in the
interaction picture with respect to the Hamiltonian of uncoupled
oscillators. Averaging the flux rates over a sufficiently long
duration of time ($\gg \Delta t$), the overall energy exchange rates
given by Eqs.~\ref{P12} and~\ref{P21} are zero. It then follows from
Eqs.~\ref{a1pa1m} and~\ref{a2pa2m} that the resonators remain at
thermal equilibrium with the bath oscillators. Furthermore, at any
given instant of time, there is no violation of energy conservation
since all energy flux rates, properly accounted for by the terms in
Eqs.~\ref{a1pa1m} and~\ref{a2pa2m}, balance each other as shown in the
schematic of Fig.~\ref{figeq}. We note that a very large frequency
mismatch ($|\omega_2-2\omega_1| \gg \gamma,\kappa$) is beyond the
scope of the present theory. It requires consideration of other
nonlinear processes such as four-wave mixing to account for the large
energy difference given by $\hbar|\omega_2-2\omega_1|$ associated with
the microscopic nonlinear mixing process. Also, realizing strong
nonlinearity in this regime is highly impractical. Therefore, we do
not discuss it here.

As a useful comparison, we further demonstrate that an alternative
semiclassical Langevin theory leads to an unreasonable deviation from
thermal equilibrium condition for the same parameters. See
Fig.~\ref{figclass}. We note that semi-classical Langevin theories can
accurately describe thermal fluctuations phenomena in the classical
regime ($\hbar\omega \ll k_B T$). In the non-classical regime
($\hbar\omega \gtrsim k_B T$), they are reliable for linearly coupled
oscillators~\cite{zhu2013temporal,karalis2015temporal} and also for
isolated nonlinear (anharmonic)
oscillators~\cite{khandekar2015radiative,dykman1975spectral}.
However, semi-classical Langevin theory fails to describe the
nonlinear system considered here. Rather than a simple
quantum--classical distinction, this failure stems from its inadequacy
in accurately capturing the complicated nonlinear mixing of thermal
energy between the oscillators of unequal non-classical frequencies
which is required to maintain thermal equilibrium of the oscillators
with the heat baths. These important and subtle theoretical details
justify the use of quantum theory of damping in the present work which
is otherwise uncommon in the field of thermal radiation. The use of
quantum theory in this field has been previously considered for
studying temporal dynamics of radiative heat transfer for
\emph{linearly} coupled plasmonic
nanosystems~\cite{biehs2013dynamical}. However, the temporal dynamics
can also be studied in the linear regime using the semi-classical
Langevin theory~\cite{zhu2013temporal,karalis2015temporal}. Since our
focus lies elsewhere, we do not discuss this separate, interesting
topic of temporal dynamics here.

\section{Far-field thermal-emission enhancement}
\label{sec:far-field}

Having established that the quantum theory provides a reasonable
description of thermal equilibrium, we now use it to analyze far-field
thermal radiation from the same resonant system under the
nonequilibrium condition of $T_d \gg T_e$. We assume temperatures
$T_d=600$K and $T_e=0$K and compute the thermal radiation power given
by Eq.~\eqref{power}. We assume the resonators to be frequency-matched
($\omega_2 =2\omega_1$) but we remark that any frequency mismatch of
the order of the linewidth negligibly affects the results described
below. Also, throughout the manuscript, the linear coupling between
the resonators is considered to be negligible because of the vast
difference in the frequencies ($\omega_2 \neq \omega_1,
|\omega_2-\omega_1| \gg \gamma_{jl}$).

\begin{figure*}[t!]
  \centering \includegraphics[width=\linewidth]{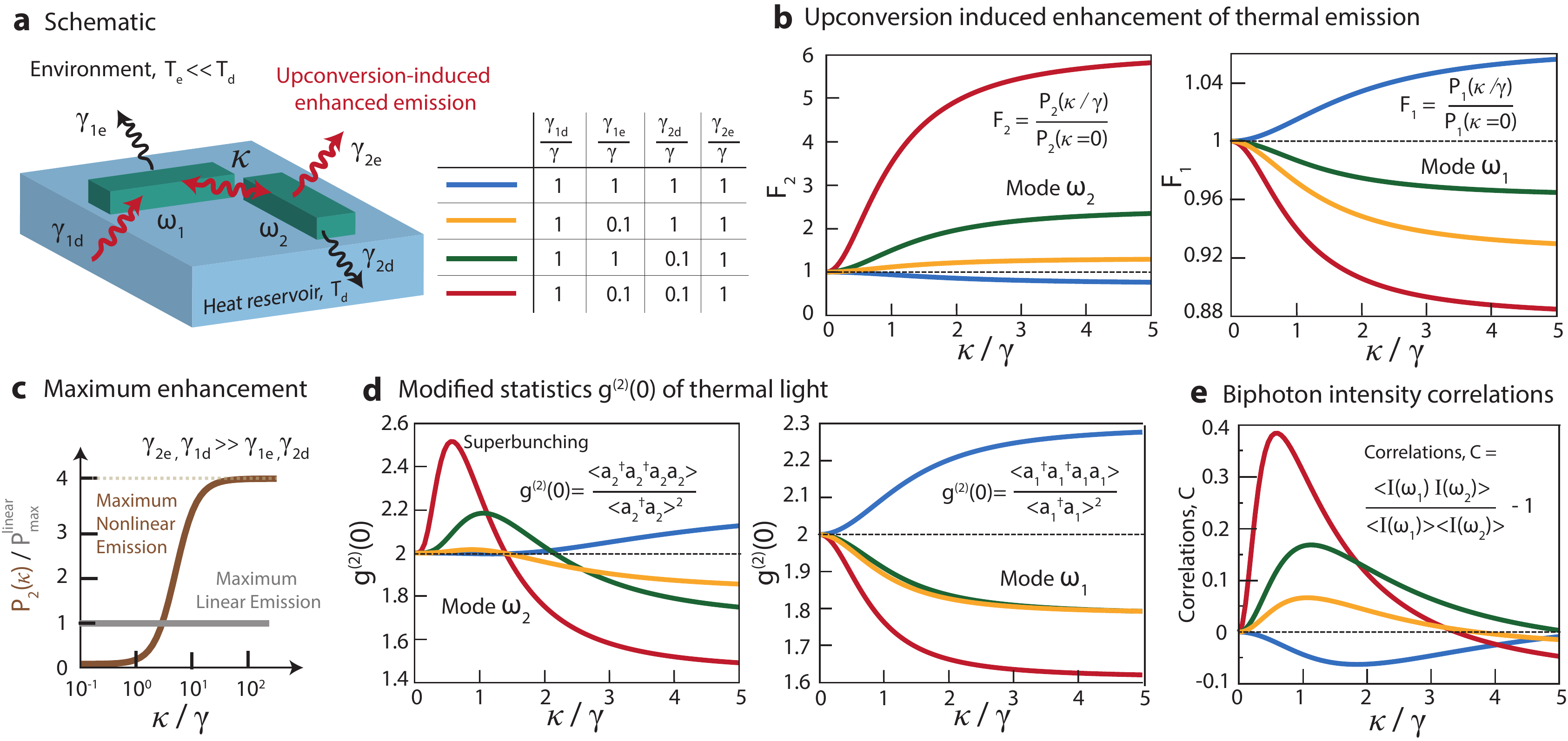}
  \caption{(a) Far-field thermal emission from a doubly resonant
    nanophotonic system at temperature $T_d=600$K into external
    environment at temperature $T_e=0$K. Various decay/coupling
    channels are shown as arrows with their directionalities
    indicating the flow of energy in the regime $T_d \gg T_e$. Red
    arrows indicate favorable channels for
    nonlinear-upconversion-induced enhancement of thermal emission at
    $\omega_2$. We consider representative normalized decay rates
    shown in the table (assuming $\gamma = 10^{-5}\omega_1$, same as
    in Fig.~\ref{figeq}) to explore thermal-radiation features in
    (b,d,e). Figures in (b) demonstrate the
    nonlinear-upconversion-induced enhancement of thermal emission at
    $\omega_2$ (left figure) and the associated suppression of thermal
    emission at $\omega_1$ (right figure), as anticipated by the
    schematic of this mechanism in Fig.~\ref{planck}. Figure(c)
    compares the maximum achievable enhancements in linear versus
    nonlinear regimes, demonstrating an enhancement factor of 4 beyond
    the standard blackbody limit for linear systems. Figure (d) shows
    that the statistics of intensity fluctuations characterized by
    $g^{(2)}(0)$ are modified at frequencies $\omega_2$ (left figures)
    and $\omega_1$ (right figures).(e) Because of the nonlinear
    coupling, the resulting thermal emission also exhibits
    correlations between the intensities [$I(\omega)$] collected at
    frequencies $\omega_1$ and $\omega_2$. The figures indicate that
    these nonlinear effects are large when the favorable channels
    $\gamma_{1d},\gamma_{2e},\kappa$ (red arrows) dominate the
    spurious channels $\gamma_{1e},\gamma_{2d}$ (black arrows).  }
  \label{figneq}
\end{figure*}

{\bf Nonlinear upconversion-induced enhancement of thermal emission:}
The schematic in figure~\ref{figneq}(a) depicts various radiative heat
exchange channels where the directionalities of the arrows indicate
plausible directions of net energy flows under the condition $T_d \gg
T_e$. Unlike the thermal equilibrium behavior discussed in the
previous section, here the upconversion and downconversion rates are
no longer balanced. Under a suitable condition, thermal energy at
$\omega_1$ can be upconverted to $\omega_2$ and emitted into the
external environment faster than its rate of downconversion back to
$\omega_1$. This imbalance provides a new mechanism of enhancing the
thermal emission at $\omega_2$. Intuitively, this mechanism is
efficient when the spurious decay rates $\gamma_{1e},\gamma_{2d}$
(shown by black arrows in the schematic) are small and favorable
decay/coupling rates $\gamma_{1d},\gamma_{2e},\kappa$ (red arrows) are
large. Accordingly, we analyze below four possible combinations of
these decay rates as shown in the table and use the same color code
for all figures in Figs.~\ref{figneq}(b,d,e).
 
Figure~\ref{figneq}(b) plots the ratio of thermal emission powers,
$F_j = P_j(\kappa)/P_j(0)$ for $j=[1,2]$, as a function of nonlinear
coupling $\kappa$. It captures the enhancement of thermal emission
beyond the linear regime at $\omega_2$ due to nonlinear upconversion
of thermal radiation at $\omega_1$. The thermal emission power in the
linear regime is
$P_2(0)=2\frac{\gamma_{2e}\gamma_{2d}}{\gamma_2}\hbar\omega_2
\nb_{\omega_2,T_d}$ where $\gamma_2=\gamma_{2e}+\gamma_{2d}$. As
evident from the red curve in Fig.~\ref{figneq}(b) corresponding to
operating regimes favorable for efficient nonlinear upconversion,
enhancement factors much greater than $1$ are realized. The additional
thermal energy in the emitted power essentially comes from the
nonlinear upconversion of thermal energy abundantly available
($\nb_{\omega_1,T} \gg \nb_{\omega_2,T}$) at low frequency
$\omega_1$. Consequently, an opposite trend at $\omega_1$ (right
figure) is expected and observed which shows the reduction of thermal
emission power due to energy depletion.

{\bf Maximum nonlinear enhancements:} We further predict the maximum
enhancement compared to linear thermal emission. Since thermal
radiation at $\omega_2$ is directly proportional to the mean photon
number $\langle n_2 \rangle$, we can maximize it by keeping only the
energy exchange channels characterized by decay rate $\gamma_{1d}$ and
the nonlinear coupling $\kappa$. Based on this intuition, we
numerically find that $\langle n_2 \rangle$ which is otherwise bounded
in the linear regime by $\nb_{2}=\gamma_{2d}
\nb_{\omega_2,T}/(\gamma_{2d}+\gamma_{2e})$ approaches
$\nb_{\omega_2,T}$ in this nonlinear regime. When $\gamma_{2e} \gg
\gamma_{2d}$, the linear thermal emission is small since $\nb_{2} \ll
\nb_{\omega_2,T}$. For this bad cavity regime, the nonlinear
upconversion of thermal energy at $\omega_1$ allows the resonator to
reach $\nb_2 = \nb_{\omega_2,T}$, enhancing thermal emission by large
factor of $(\gamma_{2e}+\gamma_{2d})/\gamma_{2d}$.

While the above comparison is for given decay rates of the resonant
system, we demonstrate another useful comparison in
Fig.~\ref{figneq}(c) between the maximum thermal emission powers
realizable in linear versus nonlinear regimes for a resonant mode of
the same total linewidth ($\gamma_2$). It is known that a perfect
linear thermal emitter satisfies the rate-matching condition
($\gamma_{2d}=\gamma_{2e}=\gamma_2/2$)~\cite{zhu2013temporal} and
emits the maximum power given by $P_2^{\text{max}}(0)=
\frac{\gamma_2}{2}\hbar\omega_2 \nb_{\omega_2,T_d}$. This linear limit
can be surpassed in the nonlinear regime where the maximum power,
$P_2^{\text{max}}(\kappa)= 2\gamma_2\hbar\omega_2 \nb_{\omega_2,T_d}$,
is obtained when the favorable channels dominate
($\gamma_{1d},\kappa,\gamma_{2e} \gg \gamma_{2d},\gamma_{1e}$). It
follows that the maximum thermal emission obtained via nonlinear
upconversion of thermal energy of a \emph{single} resonant mode is
four times larger than the maximum thermal emission power in the
linear regime. This analysis proves that the known bounds on the
enhancement of frequency-selective thermal emission can be surpassed
via nonlinear spectral redistribution of energy. This finding is
particularly important in the context of recent works on the far-field
emission enhancements from subwavelength
emitters~\cite{biehs2016revisiting,fernandez2018super,thompson2018hundred}
and the fundamental bounds on radiative heat
transfer~\cite{miller2016fundamental,buddhiraju2018thermodynamic,
  molesky2019bounds,molesky2019fundamental}.

%Of course, we have not established any tight bounds on nonlinear
%enhancements in this work which will necessarily require further
%inquiries such as enhancements feasible in multi-resonant systems,
%generalized Kirchhoff's laws~\cite{miller2017universal} and
%thermal-emission sum rules for nonlinear media.

{\bf Modified statistics and biphoton intensity correlations:} In
addition to the thermal-emission enhancements beyond the linear
regime, we find that the statistical nature of intensity fluctuations
quantified by $g^{(2)}(0)$, is also modified because of the nonlinear
mixing of thermal energy. Figure~\ref{figneq}(d) depicts the
modification in $g^{(2)}(0)$ as a function of nonlinear coupling
$\kappa$ for both resonators. Evidently, super-bunching $g^{(2)}(0) >
2$ can also be observed for finite nonlinear coupling. Furthermore,
Fig.~\ref{figneq}(e) shows that the emitted thermal radiation
collected by detectors at distinct frequencies $\omega_j$ for
$j=[1,2]$ can exhibit correlations. We quantify these correlations as,
$C = \frac{\langle I(\omega_1)I(\omega_2) \rangle}{\langle I(\omega_1)
  \rangle \langle I(\omega_2) \rangle} - 1$, where the correlations
are zero when there is no coupling between the resonators ($C=0$ when
$\kappa=0$). Here $I(\omega)$ denotes the intensity which is
proportional to $\langle a_j^{\dagger}(\omega)a_j(\omega)\rangle$. As
demonstrated, the correlations are nonzero for finite nonlinear
coupling $\kappa$ and are quite large so as to make them
experimentally noticeable. We highlight that these features are
obtained by heating the nonlinear system without using any external
signal. Also, they indicate new degrees of freedom (statistics and
biphoton correlations) for thermal-radiation engineering, which is
otherwise concerned with the control of magnitude, directionality,
spectrum and polarization of emitted
light~\cite{li2018nanophotonic,baranov2019nanophotonic}. Finally, we
note that we have checked (but not shown) that the nonlinear system
does not lead to quantum entanglement (using inseparability
criterion)~\cite{peres1996separability} and quantum
squeezing~\cite{scully1999quantum}. Other interesting questions such
as temporal dynamics from the perspective of quantum optics are not
explored since the focus is on far-field thermal-radiation which is
described in the steady state.

{\bf Potential experimental implementation:} The underlying system
considered in the present work is quite
well-known~\cite{walls2008quantum,scully1999quantum}. Furthermore, its
optimization for maximizing the nonlinear frequency mixing is
relentlessly explored by many
scientists~\cite{pernice2012second,rivoire2009second,
  bi2012high,lin2016cavity} because of its importance for
biotechnology (bio-sensing)~\cite{campagnola2003second}, material
science (spectroscopy)~\cite{heinz1982spectroscopy} and quantum
science (squeezed, entangled, single photon
sources)~\cite{walls2008quantum,scully1999quantum}. Our goal here is
to point out its potential for thermal science. To accelerate research
in that direction, we note experimental observability of these effects
by making quantitative predictions based on structures described in
the existing literature~\cite{pernice2012second,rivoire2009second,
  bi2012high,lin2016cavity}. It was shown in~\cite{lin2016cavity} that
using typical materials such as AlGaAs ($\chi^{(2)} \sim 100$pm/V) in
optimized microposts and grating structures, the nonlinear coupling
$\kappa=\beta\sqrt{\frac{\hbar\omega_1}{2}}$ where
$\beta=0.01\frac{\chi^{(2)}}{\sqrt{\epsilon_0 \lambda_1^3}}\frac{2\pi
  c}{\lambda_1}$ (derivation provided in the appendix) can be
realized. For a wavelength of $\lambda_1 = 2\mu$m, this results in the
nonlinear coupling $\kappa = 3\times 10^{-8}\omega$. Since dissipative
and radiative decay rates should be of the order of the nonlinear
coupling $\kappa$ to observe appreciable nonlinear effects, this will
require the same resonators of quality factors of $Q \sim
10^{8}$. Another recent work~\cite{lee2014giant} explores the use of
multiple-quantum-well semiconductor
hetero-structures~\cite{rosencher1996quantum} to realize orders of
magnitude larger material nonlinearities ($\chi^{(2)} \sim
10^5$pm/V). A combination of these two approaches can potentially
realize $\kappa \sim 10^{-5}\omega$ (as explored in Fig.~\ref{figeq}
and Fig.~\ref{figneq}), lowering the quality factor requirements to $Q
\sim 10^5$ to observe the predicted nonlinear effects. One possibility
to observe strong nonlinear effects with lower quality factor
resonators is to harness the large density of surface polaritonic
states ($Q \gtrsim 100$) of planar media as explored in our earlier
work~\cite{khandekar2017near}. An extension of the present work for
that geometry requires additional considerations because of many
resonant modes and therefore, it will be analyzed in our subsequent
work. We finally summarize all suitable alternatives for detecting
nonlinear thermal radiation effects in passive systems. For the
second-order nonlinear process considered here, this includes large
density of highly confined phonon polaritons in the near-field of
polaritonic media~\cite{khandekar2017near,rivera2017making},
multiple-quantum-well hetero-structures for giant nonlinearities
($\chi^{(2)} \sim 10^5$pm/V)~\cite{lee2014giant}, enhanced
nonlinearities in two-dimensional materials such as bilayer graphene
(tunable $\chi^{(2)} \sim 10^5$pm/V)~\cite{wu2012quantum}, high-Q
bound states in the
continuum~\cite{carletti2018giant,minkov2019cavity} and inverse-design
optimization~\cite{sitawarin2018inverse}. We are confident that an
optimized nonlinear system designed in the future will reveal the
predicted effects and also be useful for many other applications.

%We are currently pursuing one viable experiment to explore the
%nonlinear upconversion of near-field thermal radiation of polaritonic
%media (at $\omega_1$) using localized subwavelgnth resonators (at
%$\omega_2 \approx 2\omega_1$) in presence of quantum well
%heterostructures ($\chi^{(2)} \sim 10^5$pm/V). However, the
%theoretical analysis is challenging due to the involvement of the
%large density of resonant states. Nonetheless, our current theoretical
%work provides an important stepping stone for going forward in this
%new direction.

\section{Conclusion} \label{sec:conclusion}

We analyzed thermal radiation from a system of two \emph{nonlinearly}
coupled resonators of \emph{distinct} frequencies which also requires
careful examination of its thermal equilibrium behavior. At
frequencies ($\hbar\omega \gtrsim k_B T$), the oscillators have
unequal average thermal energies and it is not obvious whether the
coupled oscillators are at thermal equilibrium with the reservoir. Our
quantum theoretic approach describes the thermal equilibrium behavior
of the coupled oscillators reasonably well. We also note that it
cannot be adequately captured by semi-classical Langevin theory which
is otherwise useful in the linear
regime~\cite{zhu2013temporal,karalis2015temporal} and for isolated
nonlinear (anharmonic)
oscillators~\cite{khandekar2015radiative,dykman1975spectral}. These
theoretical findings are generalizable to other systems described
using such nonlinear harmonic oscillators and invite similar studies
of nonlinear thermal-fluctuations phenomena in other fields of
research e.g. thermal nonlinearities in passive (undriven) mechanical
oscillators~\cite{gieseler2013thermal}.

Most importantly, we provided a new mechanism of enhancing the
far-field thermal emission beyond the linear blackbody limits via
nonlinear upconversion.  Spectral-engineering thermal radiation using
nonlinear media is a nascent topic and the finding that it can allow
one to surpass the linear Planckian bounds is fundamentally important
in view of recent inquiries concerning super-Planckian thermal
emission from subwavelength
emitters~\cite{biehs2016revisiting,fernandez2018super,
  thompson2018hundred,greffet2018light} and bounds on radiative heat
transfer~\cite{miller2016fundamental,buddhiraju2018thermodynamic,
  molesky2019bounds,molesky2019fundamental}. While we did not
establish any tight bounds on feasible nonlinear enhancements, our
preliminary results motivate future work on thermal photonic
nonlinearities in other systems and inquiries of generalized
Kirchhoff's laws~\cite{miller2017universal} and thermal-emission sum
rules for nonlinear media. We also found that because of the nonlinear
mixing, the emitted thermal light exhibits nontrivial statistics
($g^{(2)}(0) \neq 2$) and biphoton intensity correlations. These
features are surprising from the perspective of quantum optics because
they can be observed by heating a nonlinear system without the need
for any external signal. Finally, from the perspective of theory
development in the field of thermal radiation, we note that recent
scientific efforts have extended the existing fluctuational
electrodynamic paradigm to explore non-traditional
nonreciprocal~\cite{zhu2018theory,khandekar2019thermal} and
nonequilibrium~\cite{jin2016temperature,greffet2018light,
  khandekar2019circularly} systems. Our present work aspires to go
beyond these regimes to study nonlinearities and while doing so, also
departs from the traditional, semi-classical theoretical paradigms. We
are confident that additional fundamental discoveries will be made in
this largely unexplored territory in the near future. \\

\section*{Appendix}
\appendix 
We analyze the same doubly resonant system containing a $\chi^{(2)}$
nonlinear medium and supporting resonances at frequencies $\omega_1$
and $\omega_2=2\omega_1$. We use the semi-classical Langevin theory
and show that it leads to large deviation from thermal equilibrium
condition, which is otherwise captured reasonably well using the
quantum theory of damping described in the main text.\\ 

{\bf Semi-classical theory:}
We directly introduce the Langevin (temporal coupled mode) equations
for the resonator amplitudes based on previous
work~\cite{khandekar2015radiative}. By denoting the field amplitude as
$a_j(t)$ for $j=[1,2]$ normalized such that $|a_j|^2$ denotes the mode
energy, we write the following coupled mode equations:
\begin{align}
  \label{a1c}
  \dot{a}_1 &= [i\omega_1 - \gamma_1]a_1 - i\beta_1 a_2 a_1^* +
  D_1\xi_{1d} + \sqrt{2\gamma_{1e}}\xi_{1e} \\ \dot{a}_2 &= [i\omega_2
    - \gamma_2]a_2 - i\beta_2 a_1^2 + D_2\xi_{2d} +
  \sqrt{2\gamma_{2e}}\xi_{2e} \label{a2c}
\end{align}
Here $\xi_{jl}$ for $j=[1,2]$ and $l=[d,e]$ denote the uncorrelated
white noise sources corresponding to intrinsic dissipative ($d$) heat
bath and extrinsic ($e$) environment respectively. They satisfy the
frequency correlations,
\begin{align}
\langle \xi_{jl}^*(\omega)\xi_{jl}(\omega')\rangle =
\Theta(\omega_j,T_l)\delta(\omega-\omega')
\end{align}
where $\langle \cdots \rangle$ is thermodynamic ensemble average and
$T_l$ is the temperature. The coefficients $D_j$ are the fluctuation-
dissipation (FD) relations which are to be obtained from thermodynamic
considerations as we describe further below. The nonlinear coupling
parameters $\beta_j$ for $j=[1,2]$ can be calculated using the
standard perturbation theory approach for weak
nonlinearities~\cite{bravo2007modeling,rodriguez2007chi}. They depend
on the overlap integral of the linear cavity fields $\Ev(\omega_j)$
and are given below:
\begin{align}
\beta_1 &= \frac{\omega_1}{2} \frac{\int \epsilon_0
  E_i^*(\omega_1)\chi^{(2)}_{ijk}[E_j(\omega_2)E_k^*(\omega_1)+E_j^*(\omega_1)
    E_k(\omega_2)]} {\sqrt{\int
    \frac{\partial(\epsilon\omega)}{\partial \omega}|\Ev(\omega_2)|^2}
  (\int \frac{\partial(\epsilon\omega)}{\partial \omega}
  |\Ev(\omega_1)|^2)} \nonumber \\ \beta_2 &= \frac{\omega_2}{2}
\frac{\int \epsilon_0
  E_i^*(\omega_2)\chi^{(2)}_{ijk}E_j(\omega_1)E_k(\omega_1)}
     {\sqrt{\int \frac{\partial(\epsilon\omega)}{\partial \omega}
         |\Ev(\omega_2)|^2} (\int
       \frac{\partial(\epsilon\omega)}{\partial \omega}
       |\Ev(\omega_1)|^2)}
     \label{overlap}
\end{align}
It follows that $\beta_1=\beta_2^*=\beta$ for $\omega_2=2\omega_1$
(frequency matching condition). The power flows leaving the modes via
nonlinear coupling are:
\begin{align}
  P_{1 \rightarrow 2} &= i\beta_1 \langle a_2 a_1^{*^2} \rangle -
  i\beta_1^* \langle a_2^* a_1^2 \rangle \\ P_{2 \rightarrow 1} &=
  i\beta_2 \langle a_2^* a_1^2 \rangle - i\beta_2^* \langle a_2
  a_1^{*^2} \rangle
\end{align}
By substituting $a_j = \sqrt{\hbar\omega_j} \tilde{a}_j$ for $j=[1,2]$ in above
equations so that $\langle \tilde{a}_j^* \tilde{a}_j \rangle =
\frac{\langle |a_j|^2 \rangle}{\hbar\omega_j}$ denotes the mean photon
number and then comparing with corresponding expressions in the
quantum theory (equations (7) and (8) in the main text), it follows
that the nonlinear coefficient $\beta$ is related to the nonlinear
coupling $\kappa$ through the relation
\begin{align}
  \kappa = \beta\sqrt{\frac{\hbar\omega_1}{2}}
  \label{eqkappa}
\end{align}
By making this analogy, the nonlinear coupling $\kappa$ used in the
quantum theory can be computed from the overlap integrals in
Eq.~\ref{overlap}.

\begin{figure}[t!]
  \centering \includegraphics[width=\linewidth]{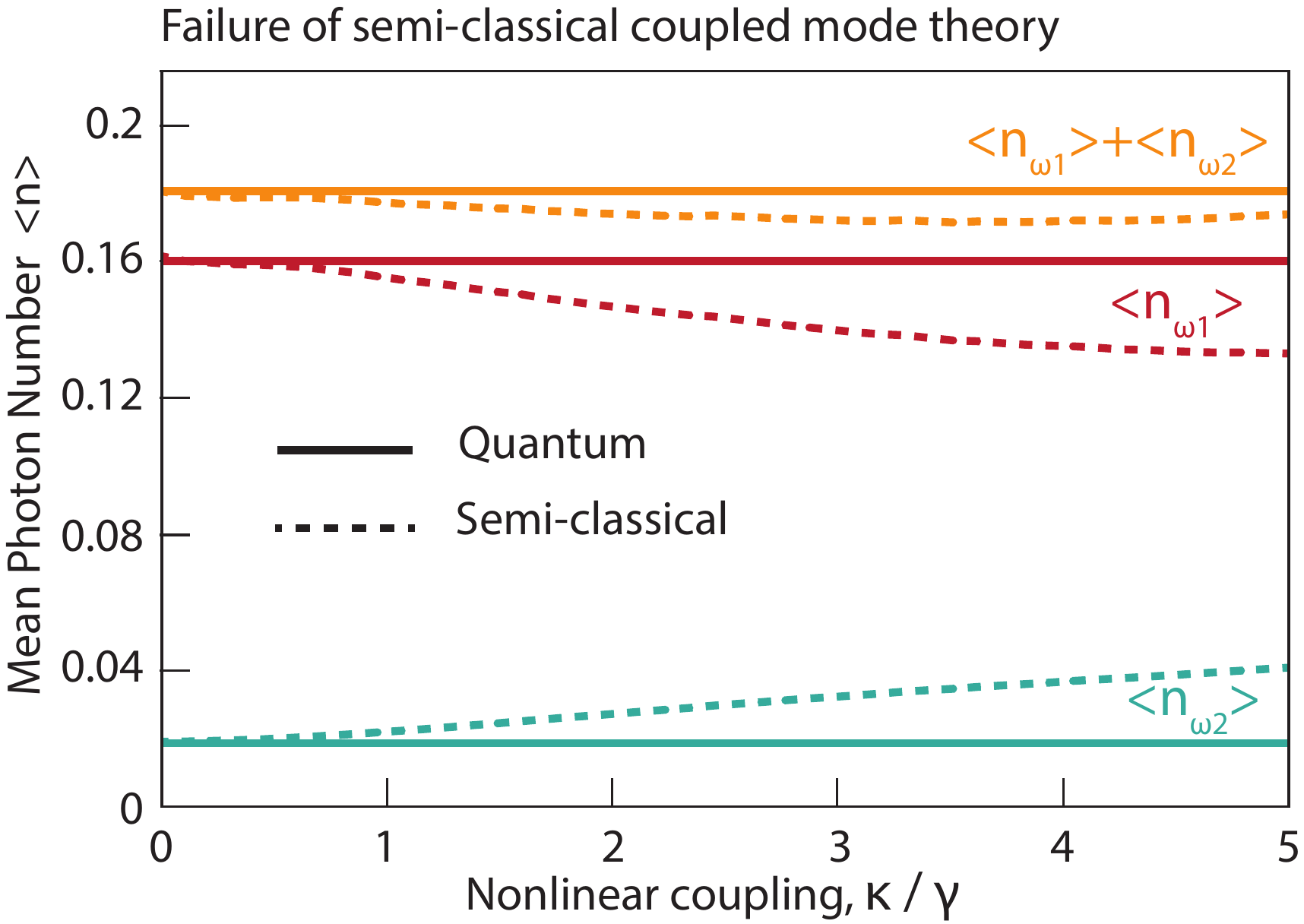}
  \caption{Steady state mean photon numbers $\langle n_j \rangle$
    obtained using semi-classical coupled mode theory are compared
    with the quantum theory for the exact same system parameters as
    analyzed in Fig.~2 of the main text. Evidently, the semi-classical
    theory leads to significant deviation from thermal equilibrium for
    finite nonlinear coupling and despite perfect frequency matching
    ($\omega_2=2\omega_1$).}
  \label{figclass}
\end{figure}

We now obtain the FD coefficients $D_j$ by transforming the stochastic
ODE into corresponding Fokker-Planck equation for the probability
distribution $P(a_1,a_1^*,a_2,a_2^*)$ which is:
\begin{align}
  \frac{dP}{dt} = \sum_{j=[1,2]} -\frac{\partial}{\partial
    a_j}K_{a_j}P - \frac{\partial}{\partial a_j^*}K_{a_j^*}P +
  \frac{1}{2}\frac{\partial^2}{\partial a_1\partial a_1^*} K_{a_1
    a_1^*}P
\end{align}
with Fokker-Planck coefficients,
\begin{align}
  K_{a_1} &=[i\omega_1-\gamma_1]a_1 - i\beta a_2 a_1^*, \hspace{20pt}
  K_{a_1^*}=K_{a_1}^* \nonumber \\ K_{a_2}&=[i\omega_2-\gamma_2]a_2 -
  i\beta^* a_1^2, \hspace{20pt} K_{a_2^*}=K_{a_2}^* \nonumber
  \\ K_{a_j a_j^*} &= K_{a_j^* a_j} = D_j^2 \Theta(\omega_j,T_d) +
  2\gamma_{je}\Theta(\omega_j,T_e), \nonumber \\ K_{a_j a_j} &=
  K_{a_j^* a_j^*}= 0, \nonumber \\
  K_{a_1 a_2} &= K_{a_2 a_1} = K_{a_1^* a_2^*} = K_{a_2^* a_1^*}=0
\end{align}
The above coefficients are derived based on the Ito interpretation of
stochastic calculus~\cite{moss1989noise,van1992stochastic}. In the
classical regime ($\hbar\omega_j \ll k_B T$), we can use the
Maxwell-Boltzmann probability distribution $P=e^{-U/k_BT}$ where
$U=|a_1|^2+|a_2|^2$ is the cavity energy. Here, the nonlinear
contributions to the cavity energy can be safely ignored since the
nonlinearity is considered perturbatively in the derivation of coupled
mode equations~\cite{khandekar2015radiative}. Using the above known
probability distribution in the classical regime, we derive the FD
coefficients $D_j=\sqrt{2\gamma_{jd}}$ where $\gamma_{jd}$ is the
corresponding dissipation rate. The same FD coefficients can be
further justified using a microscopic classical theory of damping. The
nonlinearly coupled oscillators are linearly coupled with a continuum
of classical heat bath oscillators of different effective temperatures
$k_B T_1=\Theta(\omega_1,T)$ and $k_B T_2=\Theta(\omega_2,T)$. Since
the nonlinear coupling does not affect the the system-heat bath linear
coupling in this microscopic theory, the resulting Langevin equations
have the same FD relations as in the linear regime.

We numerically simulate the stochastic equations in Eqs.~\ref{a1c} and
\ref{a2c} and compute the mean photon numbers in the steady state as
$\langle n_{\omega_j} \rangle = \frac{\langle |a_j(t)|^2
  \rangle}{\hbar\omega_j}$. Figure~\ref{figclass} compares the
semi-classical theory with quantum theory for the exact same
parameters as considered in figure 2(a) of the main text. In the
figure, we have used the relation given by Eq.~\ref{eqkappa} to
demonstrate the dependence of the mean photon numbers on the nonlinear
coupling $\kappa$ instead of the nonlinear coefficient
$\beta$. Evidently, the steady state values obtained using the
semi-classical theory indicate significant deviation from equilibrium
mean photon numbers of heat-bath oscillators as a function of the
nonlinear coupling.  Since a large relative deviation from thermal
equilibrium occurs despite perturbative nature of the nonlinearity
($\kappa \approx 10^{-5}\omega$) and the perfect frequency matching
condition $\omega_2=2\omega_1$, we conclude that the Langevin theory
fails to accurately describe this particular nonlinear
system. Intuitively, this may be expected since the resonant
frequencies are chosen in the non-classical regime ($\hbar\omega
\gtrsim k_B T$). However, it is not obvious because the semi-classical
theory is otherwise reliable in this regime for linearly coupled
resonators~\cite{zhu2013temporal,karalis2015temporal} and for isolated
single nonlinear (anharmonic)
oscillators~\cite{khandekar2015radiative,dykman1975spectral}. We
therefore conclude that the semi-classical theory fails because it
cannot accurately capture the complicated nonlinear mixing of thermal
energy between the resonant modes of non-classical frequencies which
can ensure thermal equilibrium of resonant modes with the heat
baths. Similar failures of Langevin theory applied to nonlinear
systems (as opposed to quantum versus classical distinctions) have
been noted before by van Kampen~\cite{van1992stochastic}.

We note that we have also explored several modified forms of the
semi-classical Langevin theory unsuccessfully after our prior
work~\cite{khandekar2015radiative} until now, to describe the thermal
equilibrium behavior of nonlinearly coupled resonators. One approach
involves considering multiplicative noise where the terms $D_j$ in the
coupled mode equations [Eqs.~\ref{a1c} and \ref{a2c}] are dependent on
mode amplitudes $a_j$ for $j=[1,2]$. Their unknown functional form
should be determined using the Fokker-Planck equation by enforcing the
known equilibrium probability distribution. However, the complicated
form of the multiplicative noise solution lacks any physical
justification~\cite{van1992stochastic} and also does not yield a
consistent theory. Nonetheless, we leave it as an open question
whether thermal equilibrium behavior of the nonlinear system
considered here can be reproduced reliably using a semiclassical
theory.

\section*{Funding} This work was partially supported by Defense Advanced 
Research Projects Agency under grant number N66001-17-1-4048 and the
Lillian Gilbreth Postdoctoral Fellowship program at Purdue University
(C.K.). A.W.R. was supported by the National Science Foundation under
Grant No. DMR-1454836, the Cornell Center for Materials Research MRSEC
(award no. DMR1719875), and the Defense Advanced Research Projects
Agency (DARPA) under agreement HR00111820046.

\section*{Disclosures} The authors declare no conflicts of interest.

%%%%%%%%%%%%%%%%%%%%%%% References %%%%%%%%%%%%%%%%%%%%%%%%%

%Add references with BibTeX or manually.
%\cite{Zhang:14,OSA,FORSTER2007,Dean2006,testthesis,Yelin:03,Masajada:13,codeexample}

%%%%%%%%%% If using BibTeX:
\bibliography{photon}

\end{document}